\documentclass[fleqn,usenatbib]{mnras}

\usepackage{newtxtext,newtxmath}

\usepackage[T1]{fontenc}
\usepackage{ae,aecompl}

\usepackage{graphicx}	

\title[Multi-messenger Bayesian inference of a BNS]{Multi-messenger Bayesian parameter inference of a binary neutron-star merger}

\author[M. W. Coughlin et al.]{
Michael W. Coughlin,$^{1}$\thanks{E-mail: mcoughlin@caltech.edu}
Tim Dietrich,$^{2}$
Ben Margalit$^{3,*}$
and Brian D.~Metzger$^{4}$
\\
$^{1}$Division of Physics, Math, and Astronomy, California Institute of Technology, Pasadena, CA 91125, USA\\
$^{2}$Nikhef, Science Park 105, 1098 XG Amsterdam, The Netherlands\\
$^{3}$Department of Astronomy, University of California, Berkeley, CA 94720, USA\\
$^{*}$NASA Einstein Fellow\\
$^{4}$Department of Physics and Columbia Astrophysics Laboratory, Columbia University, New York, New York 10027, USA
}

\date{Accepted XXX. Received YYY; in original form ZZZ}

\pubyear{2019}

\begin{document}
\label{firstpage}
\pagerange{\pageref{firstpage}--\pageref{lastpage}}
\maketitle

\begin{abstract}
The combined detection of a binary neutron-star merger in both gravitational waves (GWs) and electromagnetic (EM) radiation spanning the entire spectrum
-- GW170817 / AT2017gfo / GRB170817A -- 
marks a breakthrough in the field of multi-messenger astronomy. 
Between the plethora of modeling and observations, the rich synergy that exists among 
the available data sets creates a unique opportunity to constrain the binary parameters, the equation of state of supranuclear density matter, 
and the physical processes at work during the kilonova and gamma-ray burst.
We report, for the first time, Bayesian 
parameter estimation combining information from GW170817, AT2017gfo, 
GRB170817 to obtain truly multi-messenger constraints on the tidal 
deformability $\tilde{\Lambda} \in [302,860]$, total binary mass 
$M \in [2.722,2.751] M_\odot$, the radius of a $1.4$ solar mass 
neutron star $R \in [11.3,13.5] \rm km$ (with additional 
$0.2\ \rm km$ systematic uncertainty), and
an upper bound on the mass ratio of $q \leq 1.27$, all at 90\% confidence. 
Our joint novel analysis makes use of new phenomenological descriptions of the dynamical ejecta, 
debris disk mass, and remnant black hole properties, all derived from a large suite of numerical relativity simulations.  
\end{abstract}




\section{Introduction}

\begin{figure}
    \includegraphics[width=3.5in]{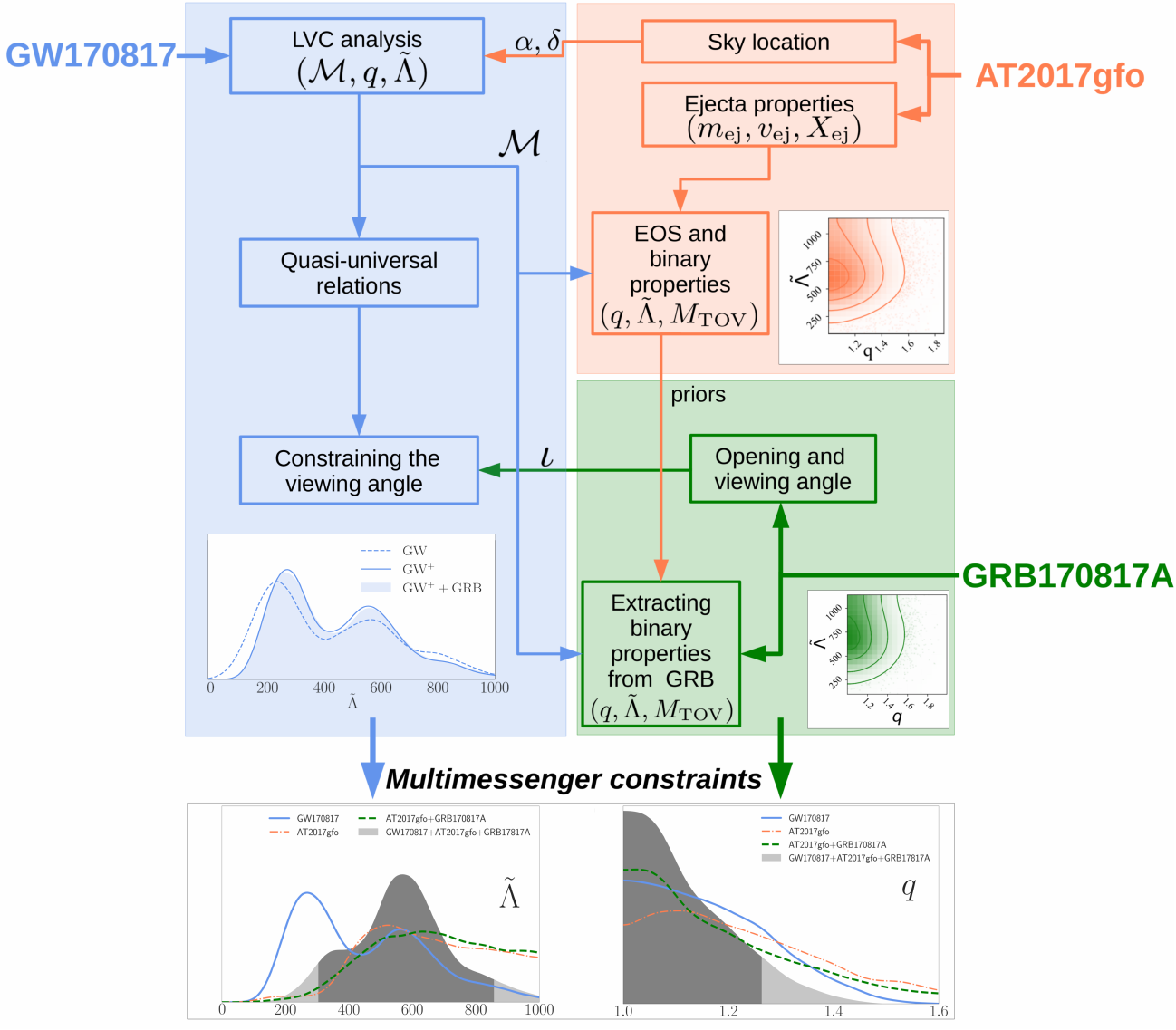}
    \caption{Flow chart of the analysis showcasing how the analysis of GW170817, AT2017gfo, and GRB170817A.
    At the bottom of the panel, we show KDE posterior distributions of the tidal deformability (left panel)
    and the mass ratio (right panel). 
    The final multi-messenger result is shown as a shaded region, where the 90\% 
    confidence interval is shaded darker. For the mass ratio, we assume a $90\%$
    upper limit and for the tidal deformability we mark the $5$ and $95$ percentiles.
    }
    \label{fig:flowchart}
\end{figure}

The combined detection of a GW event, GW170817~\citep{TheLIGOScientific:2017qsa}, 
a gamma ray burst (GRB) of short duration, GRB170817A~\citep{Monitor:2017mdv} 
accompanied by a non-thermal afterglow, and thermal emission (``kilonova'')
at optical, near-infrared, and ultraviolet wavelengths, 
AT2017gfo~\citep{GBM:2017lvd,Arcavi:2017xiz,Coulter:2017wya,Lipunov:2017dwd,
Soares-Santos:2017lru,Tanvir:2017pws,Valenti:2017ngx} from a binary neutron star (BNS) merger has enabled 
major leaps forward in several research areas.  
The latter include new limits on the equation of state (EOS) of 
cold matter at supranuclear 
densities (e.g.~\cite{De:2018uhw,Abbott:2018exr,Radice:2017lry,Radice:2018ozg,Coughlin:2018miv,
Bauswein:2017vtn,Annala:2017llu,Most:2018hfd,Ruiz:2017due,Margalit:2017dij,Rezzolla:2017aly,Shibata:2017xdx}).
One of the main goals of the nascent field of ``multi-messenger astronomy" is 
to obtain complementary observations of the same object or event. 
These observations, potentially across a variety of wavelengths and particle types, 
probe different aspects of the system.
In the case of GW170817, GW detectors such as 
LIGO and Virgo provide a highly accurate measurement of the binary chirp mass 
$\mathcal{M}=1.186M_{\odot}$, but leave the mass ratio, $q$, poorly constrained. 

A variety of studies over the last year focused on the properties of this first detection 
of a BNS system, including detailed analyses of the GW 
signal by the LVC (e.g.~\cite{TheLIGOScientific:2017qsa,Abbott:2018wiz,
Abbott:2018exr,LIGOScientific:2018mvr}) 
and external groups (e.g.~\cite{De:2018uhw,Finstad:2018wid,Dai:2018dca}), 
relying on different parameter estimation techniques and a variety of GW models. 
Despite this diversity of methods, all of the published works 
predict small tidal deformabilities, favoring relatively soft EOSs and placing upper limits on the radii of NSs. 
For this first BNS system the GW analyses broadly agree, and 
studies indicate that systematic errors are below the statistical 
errors~\citep{Abbott:2018wiz,Dudi:2018jzn,Samajdar:2018dcx}. 
However, this might not be the case for future GW observations 
with larger signal-to-noise ratios, thus emphasizing the need for further improvements in the 
current infrastructure and GW modeling.

Fortunately, deficiencies in the available GW information can sometimes be 
supplemented with EM observations, potentially improving the measurements 
of key parameters. For instance, the results of numerical relativity 
simulations were used to argue against the EOS being too soft, as the mass 
of the remnant accretion disk and its associated wind ejecta would be 
insufficient to account for the luminosity of the observed 
kilonova, e.g., \cite{Radice:2018xqa,Bauswein:2017vtn,Coughlin:2018miv}. 
Combining GW and EM observations thus provides an opportunity 
to independently constrain the binary parameters, place tighter bounds 
on the EOS, and obtain a better understanding of the physical processes and outcomes of BNS mergers. 

One of the first multi-messenger constraints on 
the tidal deformability and supranuclear EOS was
presented in \cite{Radice:2017lry}. 
Based on numerical relativity (NR) simulations, 
the authors proposed that the tidal 
deformability needs to be $\tilde{\Lambda}\geq400$ to ensure that 
a significant fraction of matter was either ejected from 
the system or contained within a debris disk around the BH remnant 
to explain the bright EM counterpart. 
Recently, \cite{Radice:2018ozg} updated this first analysis 
and obtained constraints on the tidal deformability of $\tilde{\Lambda} \in (323,776)$ and on 
the corresponding radius of a 1.4\,$M_\odot$ neutron star of 
$12.2^{+1.0}_{-0.8} \pm 0.2$\,km, performing a multi-messenger parameter 
estimation incorporating information from the disk mass \citep{Radice:2018pdn}.
To the best of our knowledge, \cite{Coughlin:2018miv} 
presented the first analysis of the lightcurves and spectra of AT2017gfo 
linking with a Bayesian analysis the kilonova properties to the source properties of the binary. 
We used the kilonova model of \cite{Kasen:2017sxr}
combined with methods of Gaussian Process Regression (GPR, \cite{Doctor:2017csx,Purrer:2014fza,Coughlin:2018miv}), 
and related a fraction of 
the ejected material to dynamical ejecta. 
Based on the analysis, 
the tidal deformability was limited to $\tilde{\Lambda}>197$. 

In addition, there have been studies placing limits on the maximum NS mass of 
a stable TOV star, $M_{\rm TOV}$.
Those studies are orthogonal to the works constraining 
the tidal deformability since both quantities $(\tilde{\Lambda},M_{\rm TOV})$
test different parts of the NS EOS. 
\cite{Margalit:2017dij} places a $90\%$ upper limit on 
the mass of a non-rotating NS of $2.17M_\odot$, 
\cite{Rezzolla:2017aly} report a maximum 
TOV mass of $2.16^{0.17}_{0.15}M_\odot$, 
and \cite{Shibata:2017xdx} provide an estimate for 
the maximum mass of $2.15-2.25M_\odot$.
All these constraints have been derived by assuming the formation of a BH after 
the merger of GW170817 and incorporating 
the measured chirp mass inferred from the GW analysis.
We employ for our analysis the maximum mass 
constraint derived in \cite{Margalit:2017dij}.
There has also been an effort to investigate the nature of GW170817, with regards to the possibility of a BNS or NSBH source, e.g. \citealt{Hinderer_2019,CoDi2019}.
For example, \citealt{Hinderer_2019} used numerical-relativity simulations and a joint analysis of GW and EM measurements to show that $<$40\% of the binary parameters consistent with the GW data are compatible with EM observations.

While overall many analyses of GW170817 and its electromagnetic signatures have been presented in the literature, 
we will present here the first to combine information from all three channels: GW170817, GRB170817A, and AT2017gfo.
Our work makes use of more available knowledge than employed in any previous 
multi-messenger analyses. In particular, our final posteriors describe the observed GW signature, 
the lightcurve data of AT2017gfo, and explain the properties of GRB170817A.
The flowchart in Fig.~\ref{fig:flowchart} highlights the interplay 
between the different observable signatures and 
presents the joint posteriors obtained on the tidal deformability $\tilde{\Lambda}$, 
the binary mass ratio $q$, and the maximum mass of a stable 
non-rotating neutron star $M_{\rm TOV}$.

\section{Analysis}

\subsection{GW170817}
We begin by analyzing GW170817 (blue shaded region of Fig.~\ref{fig:flowchart})
and use the publicly available ``low spin'' posterior 
samples (https://dcc.ligo.org/LIGO-P1800370, \cite{LIGOScientific:2018mvr}).
As these sample use the sky localization obtained from EM observations, 
they already incorporate EM information. 
Under the assumption that the merging objects are two NSs described 
by the same EOS~\citep{De:2018uhw,Abbott:2018exr}, 
we can further restrict the posterior distribution. 
For this purpose, we use the posterior samples of~\cite{Carson:2019rjx}
where a same spectral EOS representation for both stars is employed.
Finally, we discard those systems with viewing angles which are 
inconsistent with the ones obtained from the GRB 
afterglow by \cite{vanEerten:2018vgj}.

\subsection{AT2017gfo}
In the second phase of our work, we analyze the light curves of 
AT2017gfo (red shaded region in Fig.~\ref{fig:flowchart}). 
We fit the observational data~\citep{Coughlin:2018miv,Smartt:2017fuw,GBM:2017lvd} 
with the 2-component radiative transfer model of \cite{Kasen:2017sxr}. 
The usage of multiple components, proposed prior to the discovery of 
GW170817~\citep{Metzger:2014ila}, is motivated by different ejecta mechanisms contributing 
to the total $r$-process yields of BNS mergers.
The first type of mass ejection are ``dynamical ejecta" 
generated during the merger process itself. 
Dynamical ejecta are typically characterized by a low-electron fraction when 
they are created by tidal torque, but the electron fraction can extend to higher values (and thus the lanthanide abundance be reduced) in the case of 
shock-driven ejecta.  In addition to dynamical ejecta, 
disk winds driven by neutrino energy, magnetic fields, viscous evolution and/or 
nuclear recombination (e.g.~\cite{Kohri:2005tq,Surman:2005kf,Metzger:2008av,Dessart:2008zd,Fernandez:2013tya,
Perego:2014fma,Siegel:2014ita,Just:2014fka,Rezzolla:2014nva,Ciolfi:2014yla,Siegel:2017nub}) 
leads to a large quantity of ejecta, which in many cases exceeds that of the dynamical component. 
The ejecta components employed in our kilonova light curve analysis 
are related to these different physical ejecta mechanisms: 
the first ejecta component is assumed to be proportional to dynamical ejecta, 
$m_{\rm ej, 1} = \alpha^{-1} \ m_{\rm dyn}$, while 
the second ejecta component arises from the disk wind and is assumed 
to be proportional to the mass of the remnant disk, $m_{\rm ej, 2} = \zeta \ m_{\rm disk}$.  
By fitting the observed lightcurves with 
the kilonova models~\citep{Kasen:2017sxr} withing a GPR framework~\citep{Coughlin:2018miv}, 
we obtain for each component posterior distributions for
the ejecta mass $m_{\rm ej}$, the lanthanide mass fraction $X_{\rm lan}$ 
(related to the initial electron fraction), and the ejecta velocity $v_{\rm ej}$. 

The values of $m_{\rm ej}$, $X_{\rm lan}$, and $v_{\rm ej}$ obtained from our 
kilonova analysis are related to the properties of the binary and EOS using new phenomenological fits 
to numerical relativity simulations, which we briefly described below.
First, we revisit the phenomenological fit presented
in \cite{Radice:2018pdn} between the disk mass and 
tidal deformability $\tilde{\Lambda}$ to correlate the disk mass, 
$m_{\rm disk}$, to the properties of the merging binary.
Simulations following the merger aftermath suggest that the disk mass is accumulated primarily 
through radial redistribution of matter in the post-merger remnant. 
Thus, the lifetime of the remnant prior to its collapse is related to its stability 
and found to strongly correlate with the disk mass~\citep{Radice:2018xqa}.
We find that the lifetime in turn is governed to a large degree by the ratio 
of $M/M_{\rm thr}$, where $M$ is the total binary mass and $M_{\rm thr}$ is 
the threshold mass~\citep{Bauswein:2013jpa} above which the merger results in prompt (dynamical timescale) 
collapse to a black hole, 
which depends on the NS compactness and thus $\tilde{\Lambda}$. 
Therefore, $M/M_{\rm thr}$, rather than $\tilde{\Lambda}$ alone, provides a better 
measure of the stability of the post-merger remnant, and following the arguments above, 
is expected to correlate with $m_{\rm disk}$.

\begin{figure}[t]
    \centering
    \includegraphics[width=3.5in]{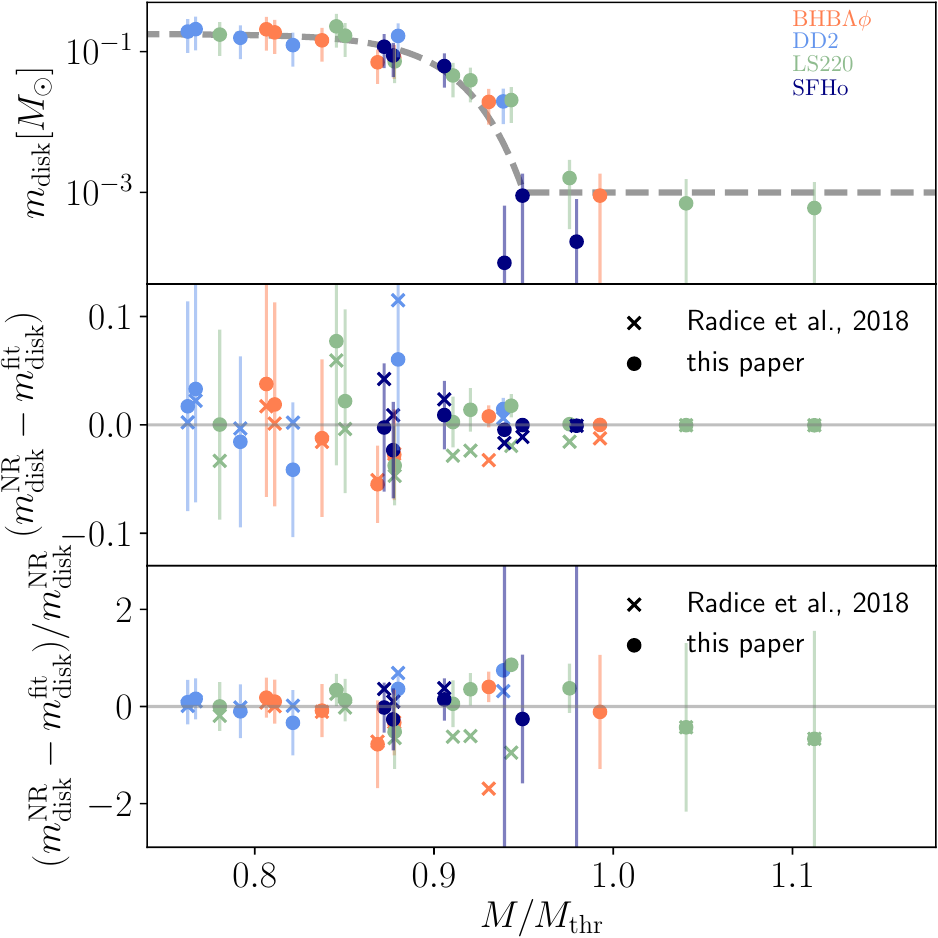}
    \caption{Disk masses as a function of the ratio between the total mass and 
    the threshold mass for prompt BH formation. 
    The disk mass estimates are obtained from the numerical relativity 
    simulations presented in \protect\cite{Radice:2018pdn}. 
    The errorbars refer to $(0.5 m_{\rm disk} + 5 \times 10^{-4} M_{\odot})$ 
    as stated in the original work of \protect\citep{Radice:2018pdn}.
    The threshold mass for prompt BH formation 
    is computed following \protect\cite{Bauswein:2013jpa}.
    We present our best fit, Eq.~\eqref{eq:mdisk_fit}, in the top panel 
    and show the absolute and fractional errors of the phenomenological fit in the middle 
    and bottom panel. We compare our results with the original version of the fit 
    presented in \protect\cite{Radice:2018pdn}.}
    \label{fig:disk_Mth}
\end{figure}

Fig.~\ref{fig:disk_Mth} shows, based on the suite of 
numerical relativity simulations of \cite{Radice:2018pdn}, 
that there indeed exists a relatively tight correlation 
between the accretion disk mass and $M/M_{\rm thr}$.
For our analysis, we will use 
\begin{small}
\begin{align}
\log_{10} &\left( m_{\rm disk} \left[ M_{\rm tot}/M_{\rm thr} \right]\right) = \nonumber \\ 
& \max \left(-3,\ a  \left(1 + b \tanh \left[ \frac{c - M_{\rm tot}/M_{\rm thr}}{d} \right] \right) \right),
\label{eq:mdisk_fit}
\end{align}
\end{small}
with $M_{\rm thr} ( M_{\rm TOV}, R_{1.6 {M_\odot}})$ as discussed in \cite{Bauswein:2013jpa} 
and the Appendix, to describe the disk mass.   
The fitting parameters of Eq.~\eqref{eq:mdisk_fit} 
are $a = -31.335$, $b=-0.9760$, $c=1.0474$, $d=0.05957$. 

Connecting the NS radius to the chirp mass and tidal deformability, 
$R = \mathcal{M} (\tilde{\Lambda}/a)^{1/6}$~\citep{De:2018uhw}, 
we conclude that the disk mass (and thus the disk wind ejecta) 
is a function of the tidal deformability, 
total binary mass, and the maximum TOV mass, 
$M_{\rm TOV}$. Therefore, both information, those on the densest portion 
of the EOS, which controls $M_{\rm TOV}$, and those from lower densities, 
as encoded in $\tilde{\Lambda}$ or $R_{1.6M\odot}$, play a role in controlling the disk 
mass and kilonova properties. 
The inclusion of these parameters and slight changes in the functional form of 
the phenomenological relation decrease the average fractional errors 
by more than a factor of 3 relative to previous disk mass estimates 
based on $\tilde{\Lambda}$ alone~\citep{Radice:2018xqa}, 
thus reducing uncertainties and errors on the EOS constraints 
obtained from kilonova observations.

Another key ingredient in our analysis is the role of the dynamical ejecta 
as the first kilonova ejecta component.  
Based on a suite of numerical relativity simulations obtained 
by different groups and codes, \cite{Dietrich:2016fpt} 
derived the first phenomenological fit for the dynamical ejecta for BNS systems. 
This fit (in its original or upgraded version) has been employed in a number of studies, 
including the analysis of GW170817~\citep{Abbott:2017wuw,Coughlin:2018miv}, 
and they have been updated in \cite{Coughlin:2018miv} and \cite{Radice:2018pdn}. 
Here, we present a further upgrade which incorporates the new numerical relativity dataset 
of \cite{Radice:2018pdn} and uses the 
fitting function of \cite{Coughlin:2018miv} (which fits 
$\log_{10} m_{\rm dyn}$ instead of $m_{\rm dyn}$). 
The extended dataset contains a total of 259 numerical relativity simulations. 
The final fitting function is 
\begin{small}
\begin{equation}
\log_{10} m_{\rm dyn}^{\rm fit} = 
\left[ a \frac{(1-2\ 	 C_1) M_1}{C_1}+b\ M_2 \left( \frac{M_1}{M_2} \right)^n +\frac{d}{2} \right] + 
[1 \leftrightarrow 2],
\label{eq:mdyn}
\end{equation}
\end{small}
with $a=-0.0719$, $b=0.2116$, $d=-2.42$, and $n=-2.905$ and $C_{1,2}$ 
denoting the compactnesses of the individual stars, a more detailed discussed 
can be found in the Appendix. 

A final ingredient in relating observational data to the binary parameters are phenomenological fits 
for the BH mass and spin. One finds that with an increasing total mass $M$, 
the final black hole mass and angular momentum increases almost linearly. 
For unequal mass mergers, $M_{\rm BH}$ and $\chi_{\rm BH}$ decrease with $M$. 
Considering the imprint of the EOS, we find that for larger values 
of $\tilde{\Lambda}$, the final black hole mass decreases, 
which follows from the observation that the 
disk mass increases with $\tilde{\Lambda}$. We finally obtain: 
\begin{small}
\begin{equation}
 M_{\rm BH} = a \left(\frac{\nu}{0.25}\right)^2 \left(M+b \ 
 \frac{\tilde{\Lambda}}{400}\right) \label{eq:MBH_fit}
\end{equation}
\end{small}
with $a=0.980$ and $b=-0.093$ and 
\begin{small}
\begin{equation}
 \chi_{\rm BH} = \tanh \left[ a \nu^2 (M+b\ \tilde{\Lambda}) +c \right] \label{eq:chi_fit} 
\end{equation}
\end{small}
with $a=0.537$, $b=-0.185$, and $c=-0.514$; 
further details are given in the Appendix.

In addition to using these fits, we use the results of \cite{Margalit:2017dij}, 
who derive a $90\%$ upper limit on the mass of a non-rotating NS of $2.17M_\odot$ based on 
energetic considerations from the GRB and kilonova which rule out a long-lived 
supramassive NS remnant, to place a prior on $M_{\rm TOV}$ between 2--2.17\,$M_\odot$. 
Combining these phenomenological relations with the lightcurve data, 
our analysis strongly favors equal or nearly equal mass systems and 
$\tilde{\Lambda} \geq 400$ (see appendix). 
We conclude that roughly $20\%$ of the first ejecta 
component is associated with dynamical ejecta, while about $20\%$ of 
the disk mass must be ejected in winds to account for the second ejecta component. 
The latter agrees with the results of long-term general relativistic 
magnetohydrodynamical simulations of the post-merger accretion disk (e.g.~\cite{Siegel:2017nub}). 
If we do not enforce constraints on $M_{\rm TOV}$, we obtain similar 
constraints in the binary parameters but with allowed values 
$M_{\rm TOV} = 2.13^{+0.35}_{-0.28}$\,$M_\odot$. This is broadly 
consistent with the results presented in~\cite{Margalit:2017dij,Rezzolla:2017aly,Ruiz:2017due}
and provides a new and independent measurement of the maximum TOV mass, which will 
become more accurate with future multi-messenger events.

\subsection{GRB170817A}
Our third and final result uses Bayesian parameter estimation of 
GRB170817A directly (green shaded region in Fig.~\ref{fig:flowchart}). 
We assume that the GRB jet is powered by the accretion of matter from the debris disk onto 
the BH~\citep{Eichler:1989ve,Paczynski:1991aq,Meszaros:1992ps,Narayan:1992iy} 
and that the jet energy is proportional to the disk mass.  Accounting for the loss of disk mass to winds, 
we connect our estimates of the disk wind ejecta from the analysis 
of AT2017gfo to the following GRB parameter estimation analysis. 
In order to assess the robustness of our conclusions, and to evaluate potential systematic uncertainties, we 
show results for three different fits to the GRB afterglow: 
\cite{vanEerten:2018vgj}, \cite{Wu:2018bxg}, 
and \cite{Wang:2018nye}. 
While the analyses of \cite{vanEerten:2018vgj,Wu:2018bxg} differ on the energy of the GRB, 
the use of either one further constrains the value of $\tilde{\Lambda}$ and 
the binary mass ratio, shifting both to slightly higher values than obtained 
through the analysis of AT2017gfo alone.

\begin{table}
  \centering
  \caption{Final multi-messenger constraints on the EOS and the 
  binary properties of GW170817. The radius constraint has to be 
  assigned with an additional $0.2\ \rm km$ uncertainty due to the employed 
  quasi-universal relations of \protect\cite{De:2018uhw}.}
  \begin{tabular}{c|c}
  \hspace*{0.3cm} Parameter \hspace*{0.3cm} & \hspace*{0.3cm}  $90\%$ confidence interval \hspace*{0.3cm}  \\
  \hline \hline
  $M$               & $[2.722,2.751] M_\odot$  \\
  $q$               & $[1.00,1.27]$  \\
  $\tilde{\Lambda}$ & $[302,860]$ \\
  $R$               & $[11.1,13.7]\ {\rm km}$ 
  \end{tabular}
 \label{tab:constraints}
\end{table}

\section{Multi-messenger constraints}
\begin{figure}
    \centering
    \includegraphics[width=3.5in]{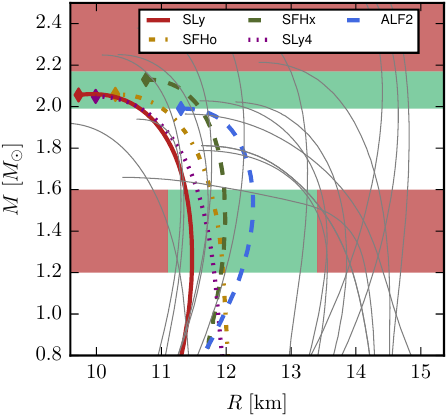}
    \caption{EOS overview including the known constraints on the maximum TOV 
    mass and the NS radius. Realistic EOSs need to fall within the green shaded regions 
    and are outside the $90\%$ confidence intervals in the red areas. }
    \label{fig:EOS}
\end{figure}

To obtain the final constraints on the EOS and binary properties, we combine the posteriors obtained 
from GW170817 and AT2017gfo+GRB170817A. The analysis of AT2017gfo and GRB170817A are highly correlated, 
as both use the same phenomenological description for the disk mass and the 
AT2017gfo posteriors are employed as priors for the GRB analysis.
However, we assume the parameter estimations results from 
the GW and EM analysis for $\tilde{\Lambda}$ and $q$ are independent from one another. 
Thus, the final multi-messenger probability density function is given by: 
\begin{equation}
 P_{\rm MMA}= P_{\rm GW170817} \times P_{\rm AT2017gfo+GRB10817A}.
\end{equation}
In principle, there are also contributions from the priors in $P_{\rm MMA}$, 
but because they are flat over the bounds considered, it is valid.
We summarize our constraints on the binary parameters and EOS in Table~\ref{tab:constraints}.
The final constraints on the tidal deformability and the mass ratio 
are shown at the bottom of Figure~\ref{fig:flowchart}, where we use 
the GRB model of \cite{vanEerten:2018vgj} (similar constraints are obtained 
with the other GRB models). 
According to our analysis, the $90\%$ confidence interval for the tidal 
deformability is $\tilde{\Lambda}\in(302,860)$. 
The distribution has its $50\%$ percentile at $\tilde{\Lambda} \sim 569$.
Relating the measured $\tilde{\Lambda}$ confidence interval to the 
NS radius~\citep{De:2018uhw}, we obtain a 
constraint on the NS radius of $R \in (11.3,13.5)\ \rm km$
(with a $\pm 0.2\ {\rm km}$ uncertainty of the quasi-universal 
relation~\citep{De:2018uhw,Radice:2018ozg} connecting $\tilde{\Lambda}$ and $R$). 
This result is in good agreement with that recently obtained by the
multi-messenger analysis presented in \cite{Radice:2018ozg}.
Considering the constraint on the mass ratio, 
we find that $q \leq 1.27$ at $90\%$ confidence. 
Combining this with the measured chirp mass, 
the total binary mass $ M =  \mathcal{M}\left( \frac{(1+q)^{2}}{q}\right)^{3/5}$ 
lies in the range $M \in [2.722,2.751]M_{\odot}$. 
The radius constraint, together with the constraint on the maximum TOV-mass, 
can be used to rule out or favor a number of proposed NS EOSs, 
as illustrated in Fig.~\ref{fig:EOS}.  

We note that there are a number of potential systematic 
uncertainties in the presented analysis, which we, however, tried to incorporate and minimize. 
In general, we have assumed that the kilonova and GRB models are sufficient 
to generate quantitative conclusions. 
To be robust against uncertainties, we have employed large systematic error 
bars for the kilonova analysis as described in \cite{Coughlin:2018miv}. 
In addition, the merger simulations and thereby the determination of ejecta 
and disk masses may still have large uncertainties because of limited resolution 
and missing physics, see e.g.~\cite{Kiuchi:2019lls}. 
The $\alpha$ and $\xi$ variables encode some of the 
uncertainty associated with this fact, as they just assume that the simulations 
are broadly correct up to a scale factor. 
In addition, while the employed GRB models are relatively simplistic, 
we have included three different GRB analyses, 
showing that they, in general, produce consistent results.

\section*{Acknowledgements}

We that Zoheyr Doctor, Michael P\"urrer, and David Radice for helpful discussions 
and comments on the manuscript. 
We are particularly thankful to 
Zack Carson, Katerina Chatziioannou, Carl-Johan Haster, Kent Yagi, Nicolas Yunes
for providing us their posterior samples analyzing GW170817 under the assumption 
of a common EOS~\citep{Carson:2019rjx}.
MWC is supported by the David and Ellen Lee Postdoctoral Fellowship at the 
California Institute of Technology. 
TD acknowledges support by the European Union's Horizon 2020 research and innovation 
program under grant agreement No 749145, BNSmergers. 
BDM is supported in part by NASA through the Astrophysics Theory Program (grant \# NNX16AB30G).
BM is supported by NASA through the NASA Hubble Fellowship grant \#HST-HF2-51412.001-A awarded by the Space Telescope Science Institute, which is operated by the Association of Universities for Research in Astronomy, Inc., for NASA, under contract NAS5-26555.




\bibliographystyle{mnras}
\bibliography{refs} 



\appendix

\section{GW170817 analysis}

\begin{figure}
    \centering
    \includegraphics[width=3.5in]{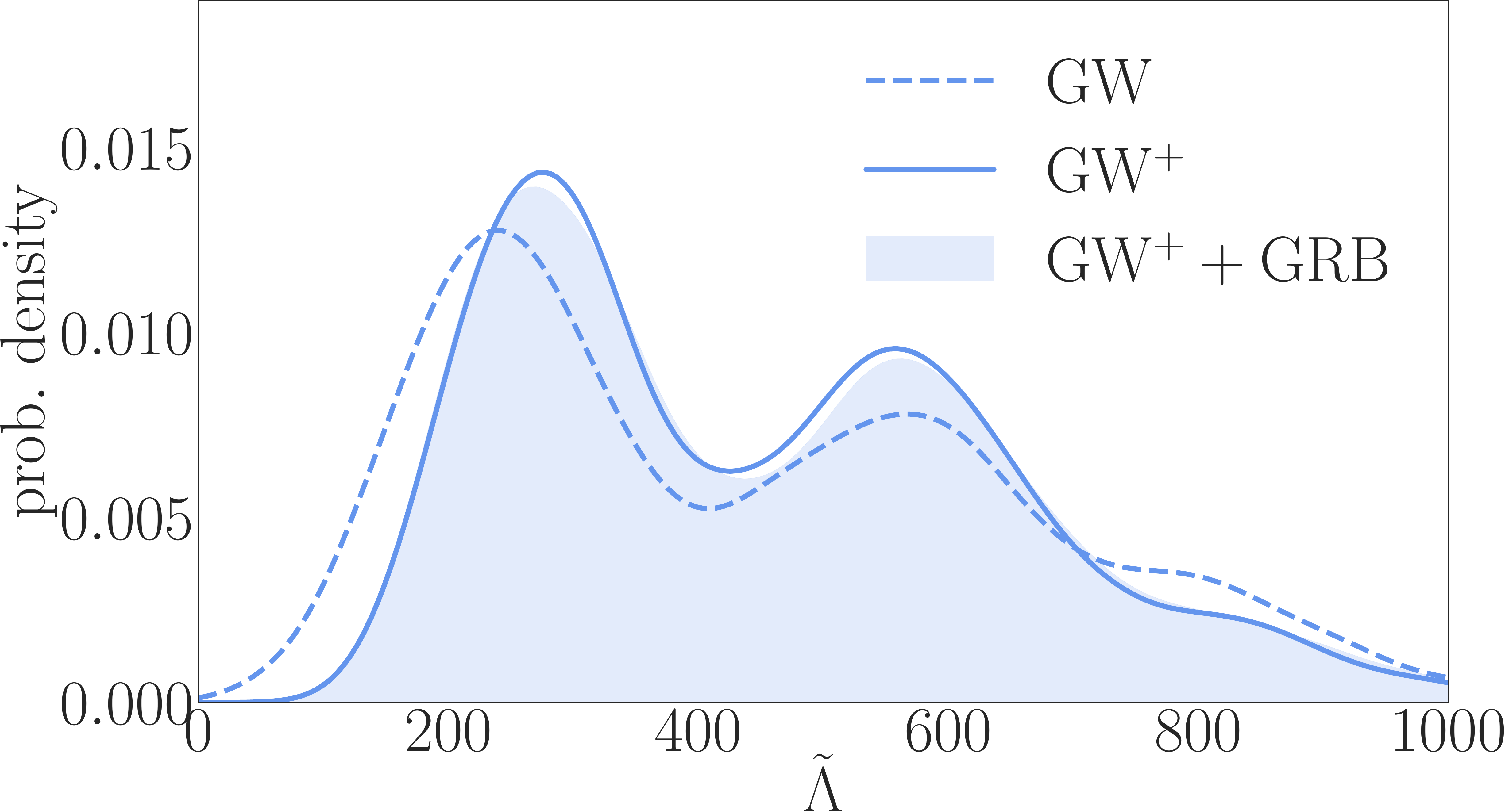}\\
    \includegraphics[width=3.5in]{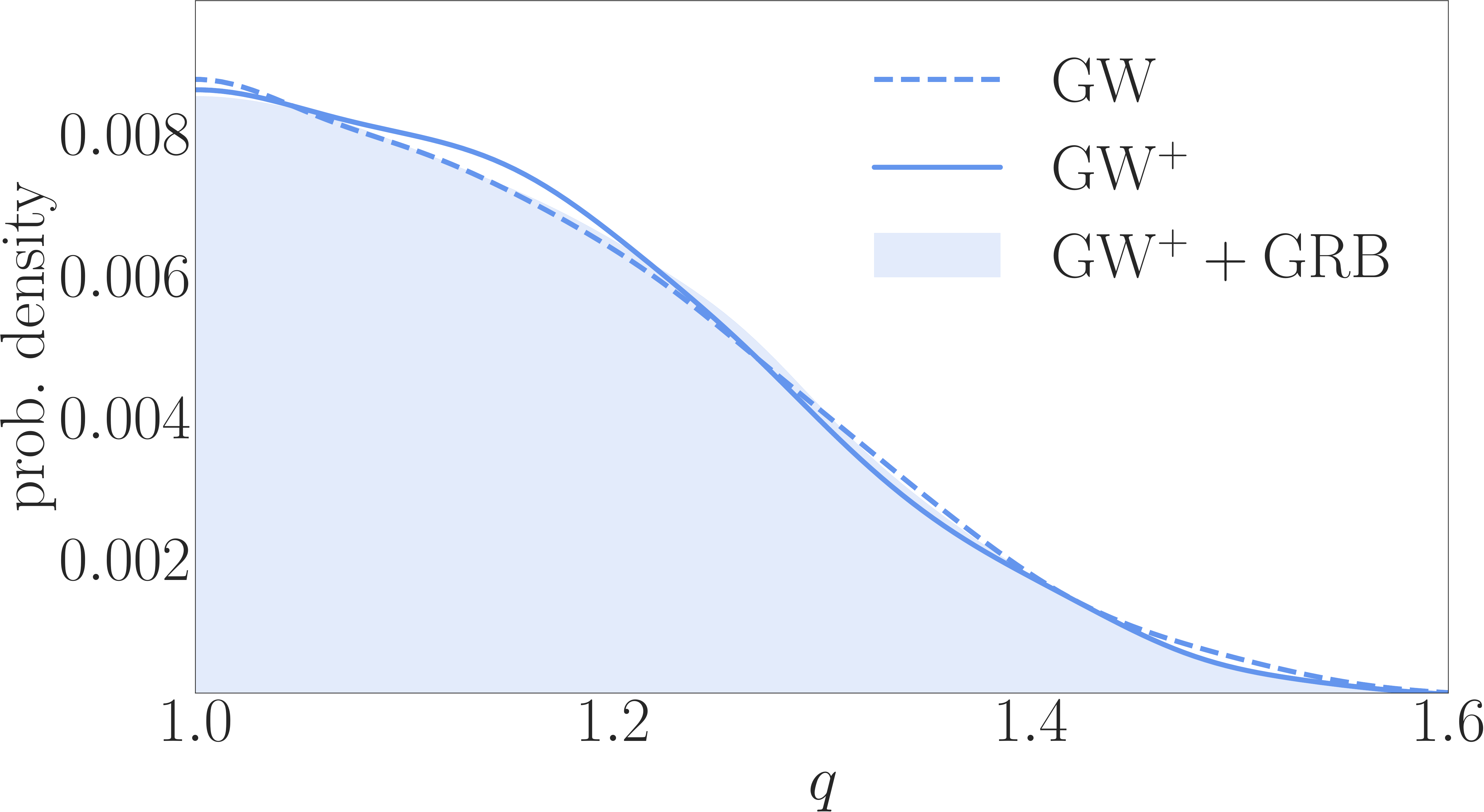}
    \caption{Probability density function for the tidal deformability (top panel)
    and the mass ratio (bottom panel)
    obtained by the analysis of GW170817. 
    The dashed line shows the original 
    posterior [denoted as `original $GW$'] 
    available at \texttt{https://dcc.ligo.org/LIGO-P1800370}, 
    the solid line shows the posterior using a spectral EOS decomposition 
    as in~\protect\cite{Carson:2019rjx}  [denoted as '$GW^+$']. 
    The shaded region marks the posterior once we incorporate the 
    viewing angle constraint and assume that the 
    merger was observed under an angle of $22\pm 12$ degree [denoted as '$GW^+ + GRB$'. 
    Notice that we use a Kernel Density Estimator (KDE) with bandwidth of $35$ 
    for the tidal deformability and $0.035$ for the mass ratio and that we
    normalize all distributions to allow a direct comparison.
    The reduced sample number due to step II and III of 
    our GW analysis (see main text) leads to larger oscillatory 
    behavior for the KDEs.}
    \label{fig:GW}
\end{figure}

We build our analysis of GW170817 on the publicly available 
posteriors released by the LVC and the results of~\citep{Carson:2019rjx}. 
We proceed in three steps: 
(i) we review the original samples, 
(ii) we restrict the analysis by ensuring that both stars are described by the same EOS,
(iii) we restrict the viewing angle based on GRB and afterglow models. 

\textbf{I. Original LIGO posteriors:}
For GW170817, we use the published results 
of the LVC~\citep{TheLIGOScientific:2017qsa,Abbott:2018wiz,Abbott:2018exr,LIGOScientific:2018mvr}. 
In particular, the publicly available 
posterior samples of~\cite{LIGOScientific:2018mvr} provide the starting 
point for our GW interpretation. 
We decide to employ the ``low spin'' assumption, which restricts the rotational frequency of individual NSs 
so that the individual dimensionless spins are restricted to 
$\chi \leq 0.05$. This restriction is motivated by the observed BNS 
in our galaxy where the fastest spinning 
NS in a BNS system (PSR J1946+2052~~\citep{Stovall:2018ouw})
will have a dimensionless spin of $\chi \sim 0.05$ at merger. 
Thus, we use the samples provided in \texttt{https://dcc.ligo.org/LIGO-P1800370}.
The analysis of the follow up LVC results~\citep{Abbott:2018wiz} improves 
over the initial results of the initial LVC results~\citep{TheLIGOScientific:2017qsa}, as a 
broader frequency band of $23-2048$ Hz, 
further improved and recalibrated detector data, 
more sophisticated waveform models~\citep{Dietrich:2018uni}, 
and the known source location from EM observations 
have been used~\citep{Monitor:2017mdv}. 

\textbf{II. Quasi-universal relations:}
While~\citep{LIGOScientific:2018mvr} did not make any assumption of the nature 
of the two merging compact objects, providing a general analysis, 
it seems natural to assume that the merging objects are 
two NSs described by the same EOS, an assumption similar 
to~\citep{De:2018uhw,Abbott:2018exr,
Radice:2018ozg}. 
Here, we employ the posterior samples of~\cite{Carson:2019rjx} 
in which a spectral EOS decomposition is employed.
As discussed in the literature, e.g.,~\cite{Chatziioannou:2018vzf,De:2018uhw,Abbott:2018wiz}, 
enforcing the two objects to be NSs described by the same EOS leads to slightly
tighter constrains on the tidal deformability (see the top panel of Fig.~\ref{fig:GW}).
We show as a dashed line the original posterior 
and the solid line marks the posterior using the spectral EOS decomposition.
Based on this result, we see that the GW data do not support large tidal deformabilities. 
This motivates that in the following analysis of the EM counterparts we will restrict 
the tidal deformabilities to $\tilde{\Lambda} \leq 1100$.

\textbf{III. The viewing angle:}
As a last step to restrict the GW posterior samples, 
we incorporate a viewing angle constraint based on the work 
of~\cite{vanEerten:2018vgj} 
(note that also other works analyzing GRB170817A and its afterglow 
can be employed and lead to similar results~\citep{Finstad:2018wid,Mooley:2017enz}).
\cite{vanEerten:2018vgj} finds that the viewing angle of the GRB170817A 
was $22 \pm 6$ degrees. For a conservative estimate, we assume $22 \pm 12$ 
degrees to allow for additional uncertainties. 
All posterior samples with viewing angles that do not fall 
inside this interval are discarded. 
The final result is shown as the shaded region 
in Fig.~\ref{fig:GW}. 

\section{Analyzing AT2017gfo}

\begin{figure}
    \centering
    \includegraphics[width=3.5in]{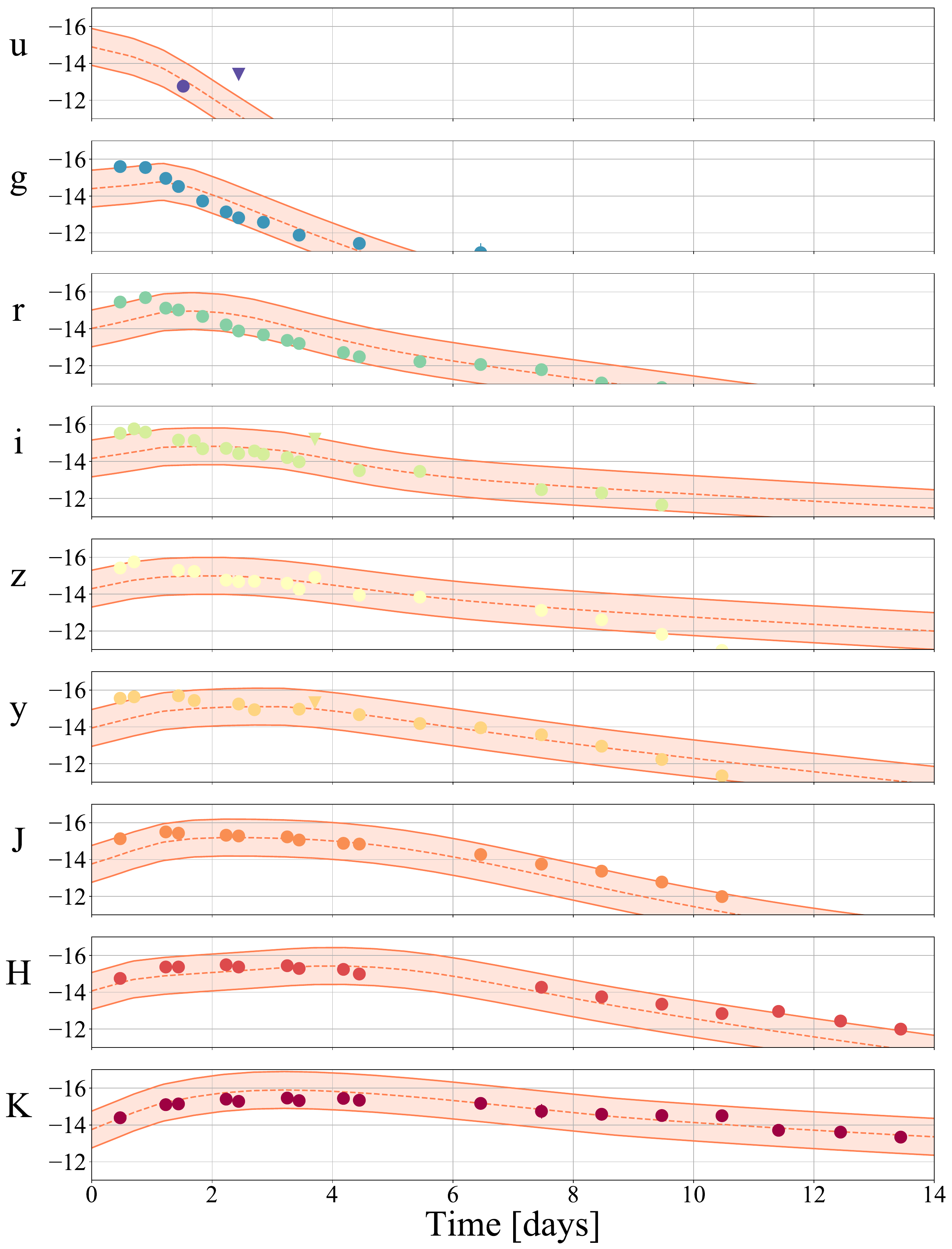}
    \caption{Observational data of AT2017gfo together with the best 
    according to our analysis and the 
    employed kilonova model.}
    \label{fig:lightcurve}
\end{figure}

Similar to the analysis of GW170817, we proceed in multiple 
steps to obtain a posterior distribution of the binary properties 
of AT2017gfo. For this purpose, we follow and extend the discussion 
in~\cite{Coughlin:2018miv} to which we refer for further details
and extensive checks of the underlying algorithms. 

\textbf{I. Modeling AT2017gfo:}
The observational data~\citep{Coughlin:2018miv,Smartt:2017fuw,GBM:2017lvd} 
(see Fig.~\ref{fig:lightcurve})
are fit with the radiative transfer model of~\cite{Kasen:2017sxr}. 
The model employs a 
multi-dimensional Monte Carlo code to solve the 
multi-wavelength radiation transport equation 
for an expanding medium. The model allows for the usage 
of a two component ejecta description. 
Each component depends parametrically 
on the ejecta mass $m_{\rm ej}$, 
the mass fraction of lanthanides 
$X_{\rm lan}$, and the ejecta velocity $v_{\rm ej}$.
These individual parameters will depend on the merger 
process and the binary parameters. 
The usage of at least two components is motivated by the presence of 
different ejecta types. Among the biggest drawbacks of 
our analysis is the assumption that both components are 
treated spherically symmetric with a uniform composition. 
Neglecting mixing of different ejecta types~\citep{Rosswog:2016dhy},
we add the two separate model components. 
We have tested the recoveries of the non-spherical models presented 
in~\cite{Kasen:2017sxr} using the spherical model.
Based on these data, there will be a viewing angle bias in the parameter recoveries here depending on the degree of asymmetry in the ejecta.

We employ a grid with ejecta masses 
$m_{\rm ej} [M_\odot]$ = 0.001, 0.0025, 0.005, 0.0075, 0.01, 0.25, 0.05, and 0.1, ejecta velocities 
$v_{\rm ej} [c]$ = 0.03, 0.05, 0.1, 0.2, and 0.3, and mass fraction of lanthanides 
$X_{\rm lan}$ = 0, $10^{-5}$, $10^{-4}$, $10^{-3}$, 
$10^{-2}$, and $10^{-1}$.
In order to draw inferences about generic sources not corresponding to one of these gridpoints, 
we adapt the approach outlined in~\cite{Doctor:2017csx,Purrer:2014fza}, 
where GPR is employed to interpolate principal components of gravitational waveforms.
The reliability of the method has been tested in~\cite{Coughlin:2018miv}.

\begin{table*}
  \centering
  \caption{Prior choices in the analysis. 
  Intervals indicate a uniform prior, 
  while $\pm$ indicates a Gaussian prior. 
  For the sGRB analyses, we draw the parameters $\tilde{\Lambda}$, $q$, $M_{\rm TOV}$, and $\zeta$ consistent with distributions found from the kilonova analysis.}
  \begin{tabular}{cc|cc|cc|cc}
\multicolumn{2}{c}{AT2017gfo} & \multicolumn{2}{c}{GRB170817A-~\citep{vanEerten:2018vgj}} & 
\multicolumn{2}{c}{GRB170817A-~\citep{Wu:2018bxg}} &  \multicolumn{2}{c}{GRB170817A-~\citep{Wang:2018nye}} \\
parameter & prior & parameter & prior & parameter & prior & parameter & prior \\
     \hline
$\tilde{\Lambda}$ & [0,1100] & 
$\tilde{\Lambda}$ & KN analysis  & 
$\tilde{\Lambda}$ & KN analysis  & 
$\tilde{\Lambda}$ & KN analysis  \\
$q$ & [1,2] & 
$q$ & KN analysis  & 
$q$ & KN analysis  & 
$q$ & KN analysis \\
$M_{\rm TOV}$ & [2.0,2.17] & 
$M_{\rm TOV}$ & KN analysis  & 
$M_{\rm TOV}$ & KN analysis  & 
$M_{\rm TOV}$ & KN analysis  \\
$\log_{10} \alpha$ & $[-2,0]$ & 
$\log_{10} E_0$ & 50.30 $\pm$ 0.84 & 
$\log_{10} E_{GRB, 50}$ & $-$0.81 $\pm$ 0.39 
& $\log_{10} \sigma$ & $[-4,-1]$\\
$\zeta$ & $[0,0.5]$ & 
$\log_{10} \epsilon$ & $[-20,0]$ & 
$\log_{10} \epsilon$ & $[-20,0]$  & 
$\zeta$ & KN analysis \\
& & $\zeta$ & KN analysis & $\zeta$ & KN analysis  \\ 
  \end{tabular}
 \label{tab:prior}
\end{table*}

\begin{figure*}
    \centering
    \includegraphics[width=3.4in]{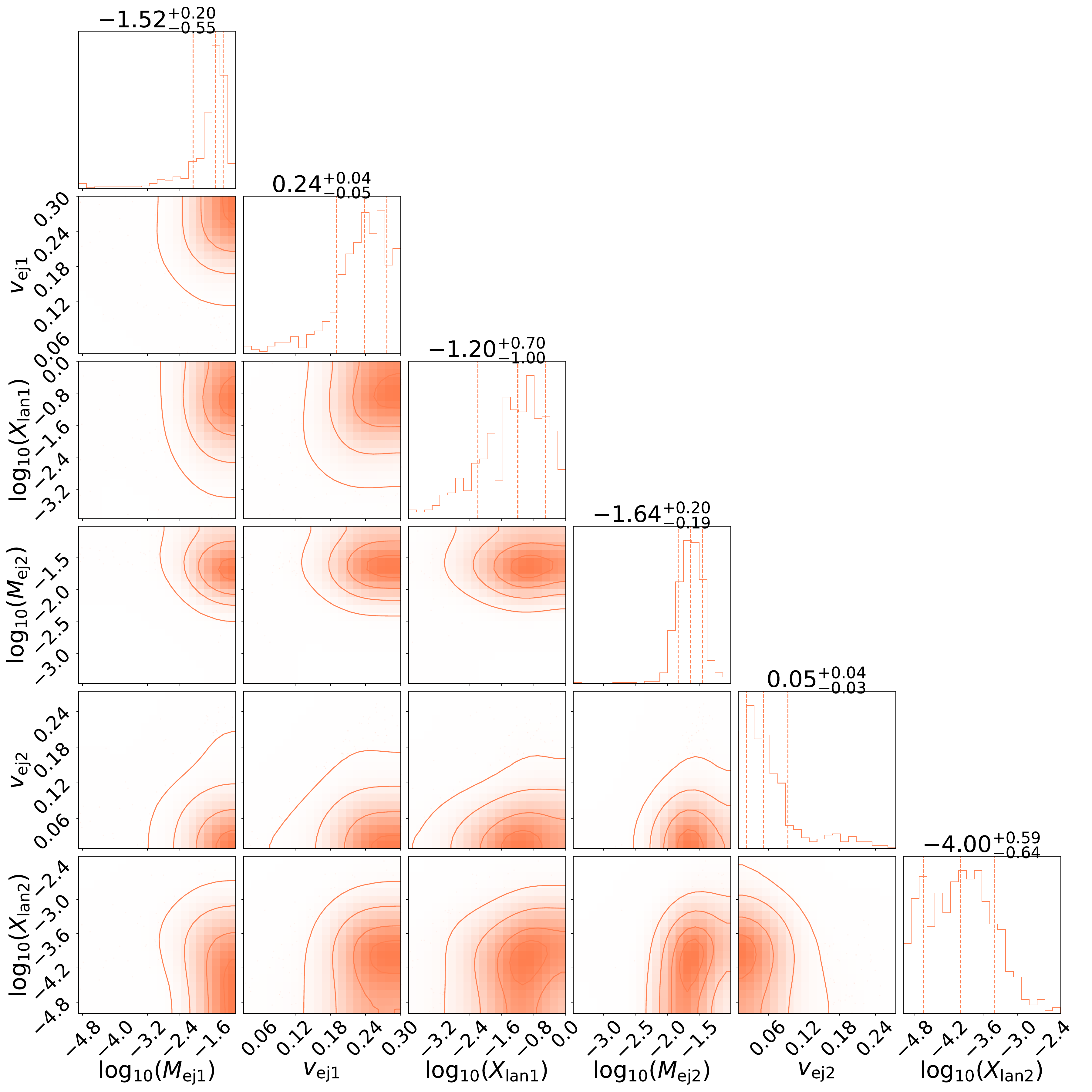} 
    \includegraphics[width=3.4in]{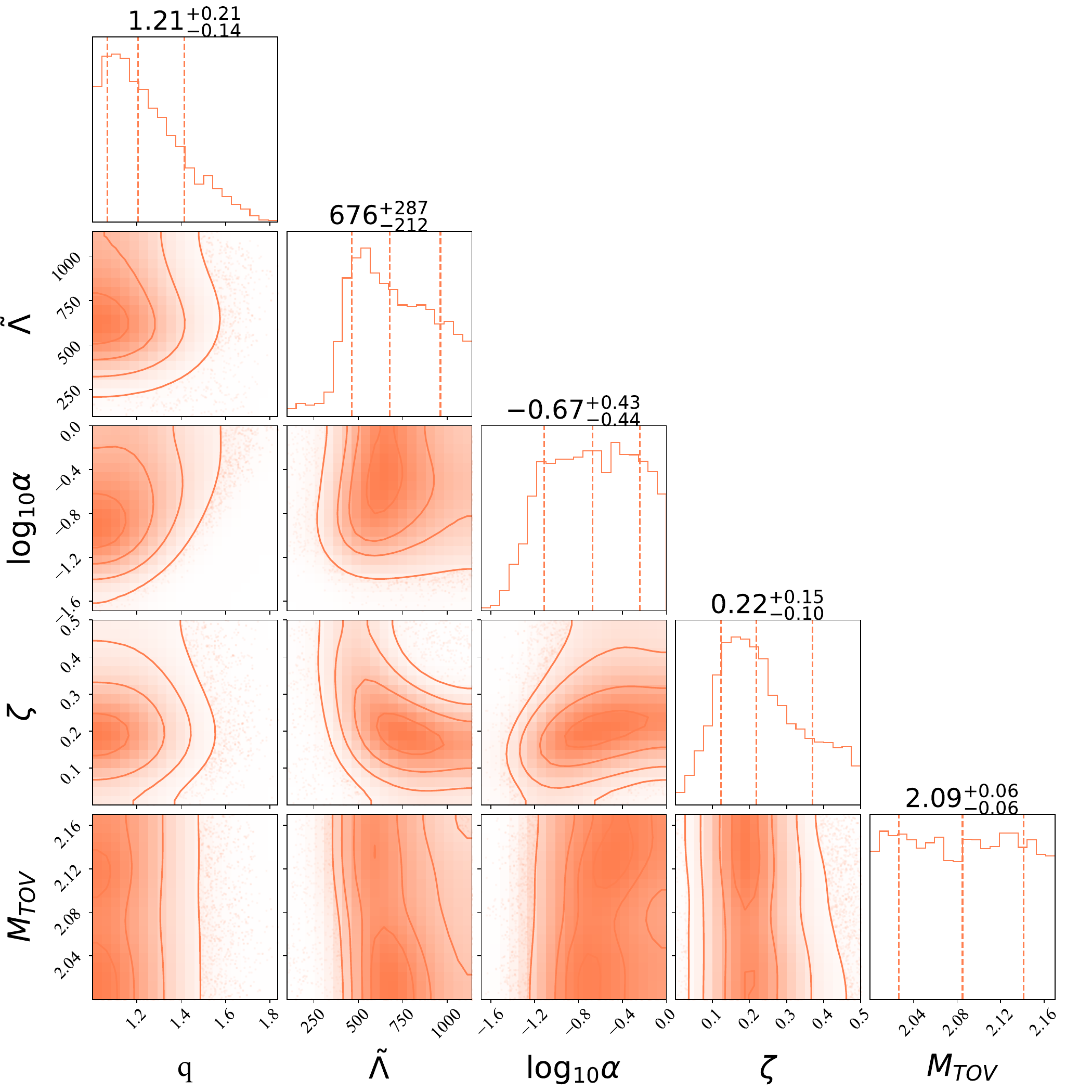}
    \caption{On the left is the posterior distribution of the ejecta properties 
    fitting the observational data
    presented in Fig.~\ref{fig:lightcurve}. 
    The shown quantities refer to the ejecta mass, velocity, and lanthanide fraction 
    of the first and second ejecta component. On the right is the posterior distribution for our analysis of AT2017gfo. 
    We present posteriors for the mass ratio $q$, the tidal deformability 
    $\tilde{\Lambda}$, the fraction of the first ejecta component 
    related to dynamical ejecta $\alpha$, the fraction of the disk mass ejected 
    as the second component ejecta, and the TOV mass $M_{\rm TOV}$.}
    \label{fig:ejecta}
\end{figure*}

For completeness, we present the lightcurves together with the observational data in 
Fig.~\ref{fig:lightcurve}. The posterior for the ejecta properties is shown 
on the left of Fig.~\ref{fig:ejecta}. 
The priors for the analysis are given in Table~\ref{tab:prior}.
We find that we are able to fit the observational data within the assumed $1$ magnitude
uncertainty~\citep{Coughlin:2018miv}. 
We want to point out that although we are able to 
fit and describe the general X-shooter 
spectra~\citep{Smartt:2017fuw,Pian:2017gtc}, the current model is 
unable to accurately represent observed wavelength specific features 
(see the discussion in~\cite{Coughlin:2018miv}).

\textbf{II. Relating ejecta properties to the binary parameters:}

To connect the individual ejecta components 
to the different ejecta mechanisms, we assume that the first ejecta component is 
proportional to dynamical ejecta, i.e., it gets released during the merger process 
and is proportional to $m_{\rm dyn}$. 
The second ejecta component arises from disk winds.
We find that constraints on the mass ratio mostly follow from this first assumption, 
and constraints on the tidal deformability arise mainly from the second ejecta component. 
While the analysis has velocity and lanthanide fraction priors 
to make these components physically motivated, 
in the case where these assumptions are incorrect, 
the analysis will break down.

With the uncertainty in the modeling of ejecta in numerical relativity simulations
and the potential systematic biases due to the missing input physics, 
we only assume that the dynamical ejecta describes a fraction of the total first component:  
\begin{equation}
    m_{\rm ej, 1} = \frac{1}{\alpha} \  m_{\rm dyn}, \qquad v_{\rm ej, 1} = v_{\rm dyn}.
\label{eq:dyn}
\end{equation}
To allow for a direct comparison with the GW analysis, we express $m_{\rm dyn}$
in terms of $\tilde{\Lambda}$. This can be achieved by writing 
the compactnesses of the individual stars as 
$C_{1,2} = 0.371 - 0.0391 \log(\Lambda_{1,2}) + 0.001056 \log(\Lambda_{1,2})^2$ 
~\citep{Maselli:2013mva,Yagi:2016bkt}, 
employing again $\Lambda_2 = q^6 \Lambda_1$, and using the definition of the tidal deformability
\begin{equation}
\tilde{\Lambda} = \frac{16}{13} \frac{\Lambda_2+\Lambda_1 q^5 + 12 \Lambda_1 q^4 + 12 \Lambda_2 q}{(1+q)^5}.
\end{equation}

The second ejecta component is related to ejecta arising from disk winds. 
Long-term simulations find that 
about $\sim 10-40\%$ ~\citep{Dessart:2008zd,Metzger:2008av,Metzger:2008jt,Lee:2009uc,
Fernandez:2013tya,Siegel:2014ita,Just:2014fka,Metzger:2014ila,
Perego:2014fma,Martin:2015hxa,Wu:2016pnw,Siegel:2017nub,
Lippuner:2017bfm,Fujibayashi:2017xsz,Fujibayashi:2017puw,
Siegel:2017jug,Metzger:2018uni,Radice:2018xqa}
of the overall disk mass can be ejected. 
Thus, it seems plausible to assume  
\begin{equation}
    m_{\rm ej, 2} = \zeta \ m_{\rm disk}, 
\label{eq:wind}
\end{equation}
i.e., the disk wind ejecta are overall proportional to the 
mass of the debris disk surrounding the remnant BH. 
Knowing that a large fraction of the disk falls into the BH 
directly after BH formation and that not all matter gets ejected, 
we restrict $\zeta$ 
to lie within $\zeta \in [0,0.5]$. 

The right part of Fig.~\ref{fig:ejecta} shows the findings of our AT2017gfo analysis, 
which we shortly summarize below: 
(i) our study favors equal or nearly equal mass systems, 
where the constraint on the mass ratio mainly arises from the 
correlation between the first component ejecta and the dynamical ejecta. 
(ii) $\tilde{\Lambda}$ shows a clear jump at $\tilde{\Lambda} \approx 400$. 
This constraint arises mainly from the second component ejecta and is related 
to the disk mass increase for larger values of $\tilde{\Lambda}$. 
(iii) Only about $20\%$ of the first ejecta component is associated to 
dynamical ejecta. 
(iv) About $22\%$ of the disk mass has to be ejected to account for the 
second ejecta component, which agrees with the disk wind ejecta found 
in long term numerical relativity simulations. 
(v) The analysis shows no strong constraint 
on the maximum allowed TOV mass. \\

\section{Analyzing GRB170817A}

In addition to including information about the viewing angle 
to restrict the GW posteriors (Fig.~\ref{fig:GW}), 
we will also present a Bayesian parameter estimation for GRB170817A directly. 
We note that a GRB-GW Bayesian approach was also suggested in \cite{Fan:2017rkg}.

To relate the GRB properties to the properties of the binary, 
we employ the typical assumption that the GRB is driven by the accretion 
of matter from the debris disk onto 
the BH~\citep{Eichler:1989ve,Paczynski:1991aq,Meszaros:1992ps,
Narayan:1992iy,Meszaros:2006rc,Lee:2007js,Giacomazzo:2012zt,Ascenzi:2018mwp} 
and therefore the energy is proportional to the disk rest mass, i.e.,  
\begin{equation} 
    E_{jet} \propto \ m_{\rm{disk}}.  \label{eq:EGRB_0}
\end{equation}
We note that based on our previous discussion, 
a fraction of the disk is ejected by disk winds. 
This part of the original disk cannot drive the GRB, so we set 
\begin{equation}
 E_{jet} = \varepsilon ( m_{\rm disk} - m_{ej,2} ) = \varepsilon m_{\rm disk} (1 - \zeta).
 \label{eq:EGRB}
\end{equation}
To connect the GRB and kilonova analysis, we reuse the $\zeta$ posterior obtained 
in the previous subsection. Similarly, we also employ the posterior distributions 
of $\tilde{\Lambda},q,M_{\rm TOV}$ as priors for our future GRB parameter estimation analysis.

We now briefly describe the three different GRB models/descriptions~\citep{vanEerten:2018vgj,Wu:2018bxg,Wang:2018nye}
used in this work.

\begin{figure*}
    \centering
    \includegraphics[width=2.2in] {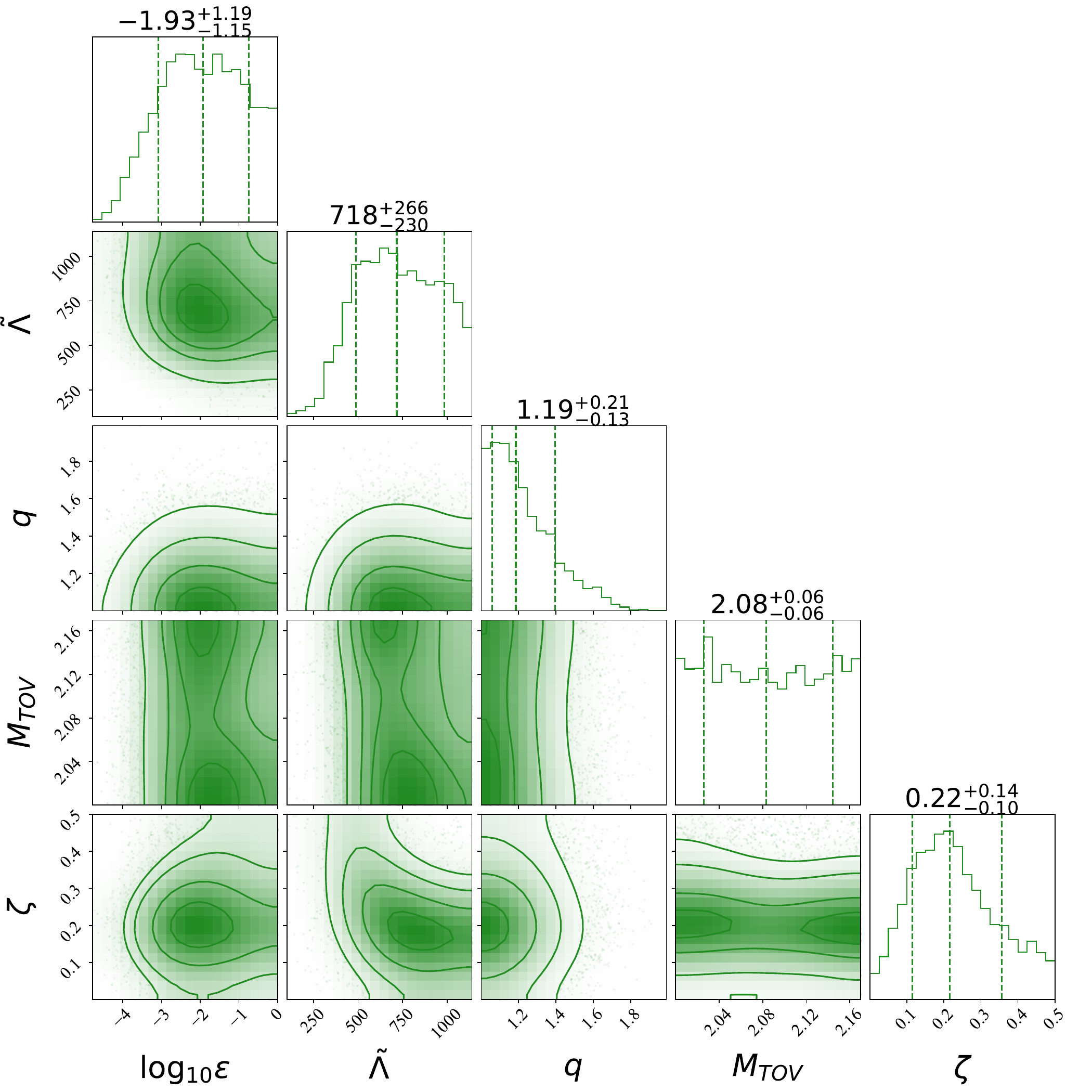}
    \includegraphics[width=2.2in]{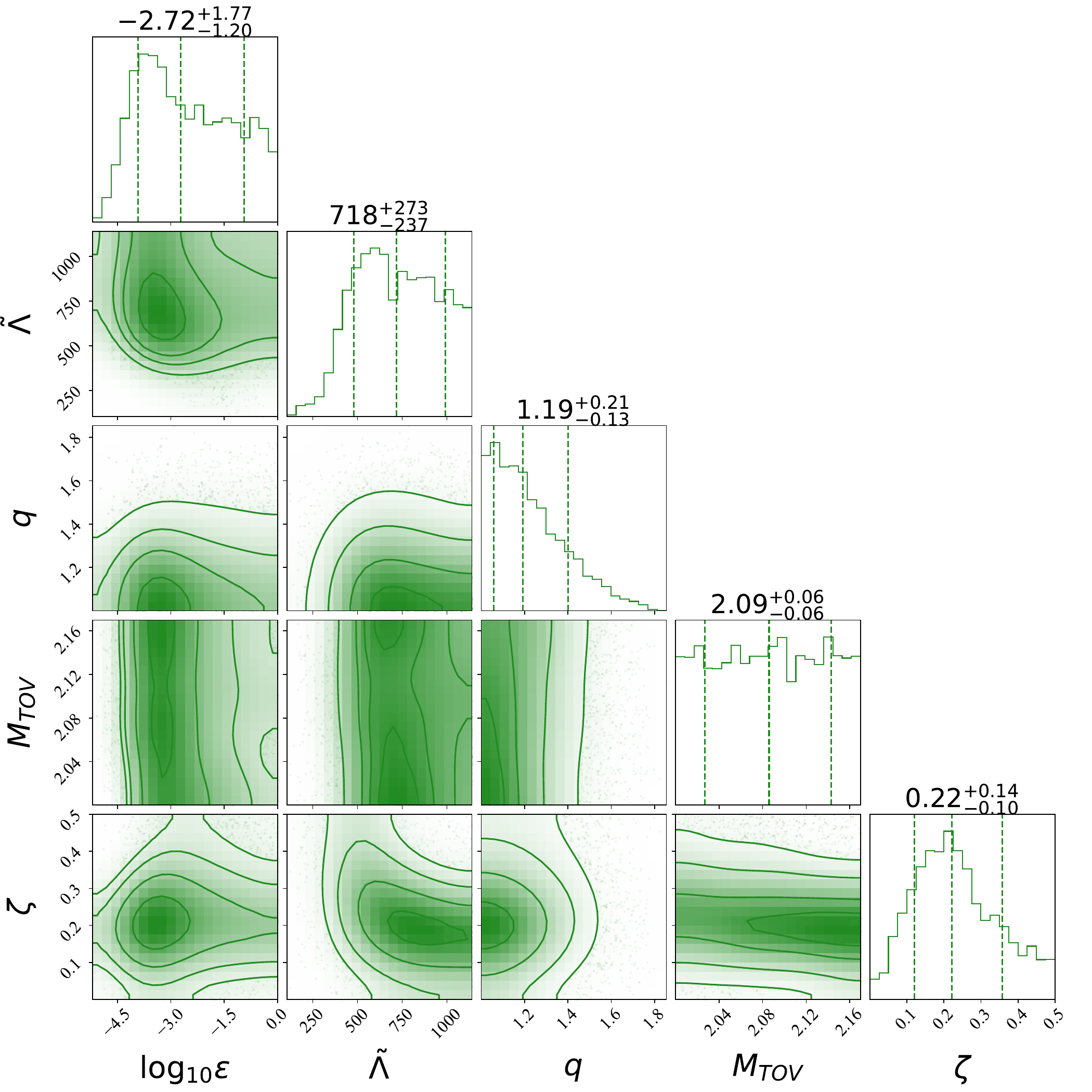}
    \includegraphics[width=2.2in]{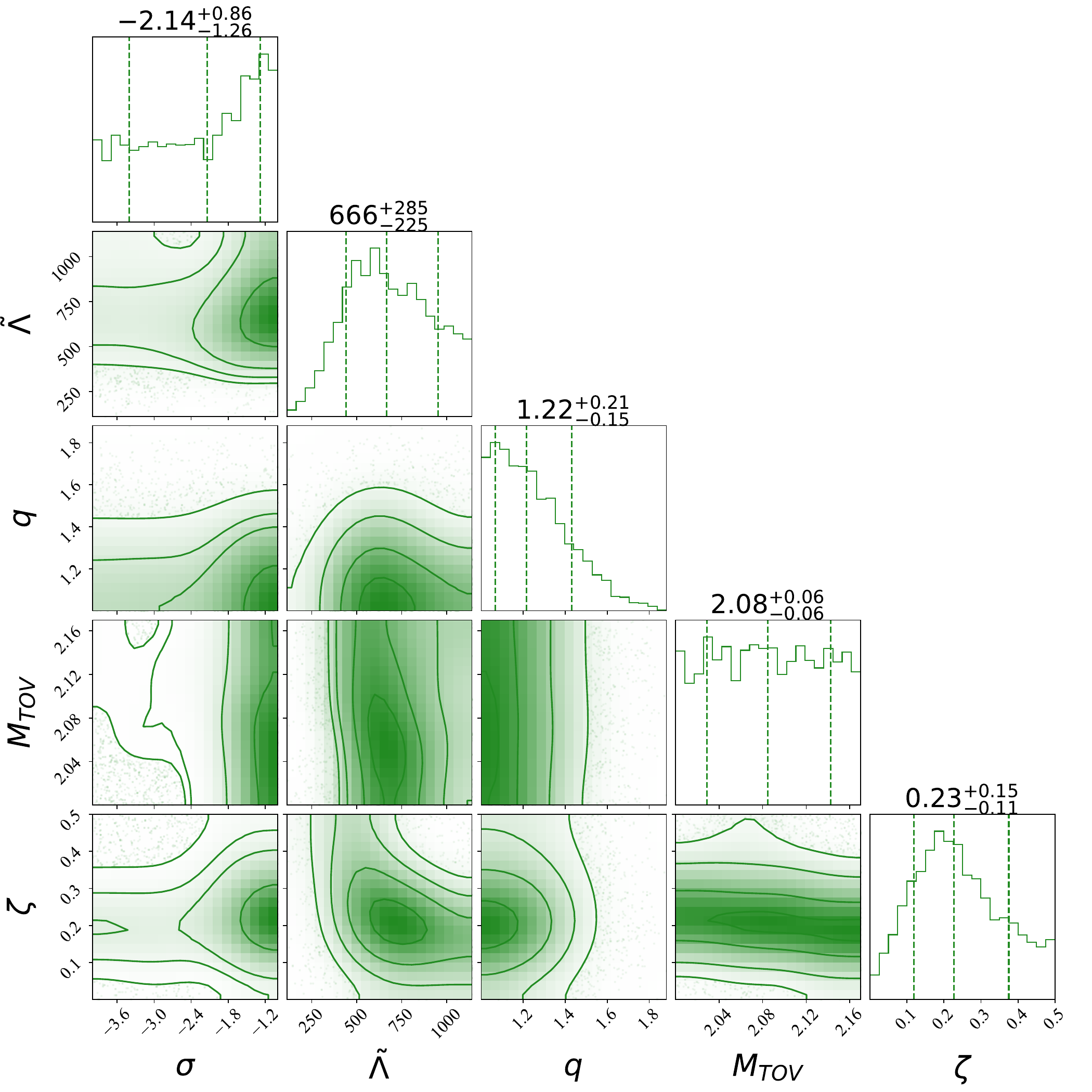}
    \caption{On the left are the posterior distributions for the GRB 
    analysis based on van Eerten et al., 2018. 
    We present constraints on the fraction of the 
    rest mass density of the disk 
    converted to trigger the sGRB, 
    the tidal deformability, the mass ratio 
    and the maximum TOV mass. In the middle are the posterior distributions for the GRB analysis using the 
    GRB energy estimated from Wu \& MacFadyen, 2018. 
    On the right are posterior distributions for the GRB analysis using the 
    GRB energy estimated from Wang et al., 2018.}
    \label{fig:GRB}
\end{figure*}

\textbf{I. The structured GRB-jet model of van Eerten et al~\citep{vanEerten:2018vgj}:}

\cite{vanEerten:2018vgj} show that the latest observations of 
the GRB170817A afterglow is consistent with the emergence of a 
relativistic structured jet (with a jet width $\theta_c$ of $4$ degrees) seen 
at an angle of $\theta_j \sim 22\pm 6$ degree from the jet axis.
This structured jet model fits well within 
the range of properties of cosmological sGRBs.
Incorporating the Gaussian structured form of the jet as proposed 
in~\cite{vanEerten:2018vgj}, the final GRB energy arising from the measured 
isotropic energy is given by
\begin{equation}
    E_{\rm jet} = e^{2\theta_c/\theta_j} E_{\rm iso}. \label{eq:EGRB_Troja}
\end{equation}
According to the analysis of~\cite{vanEerten:2018vgj}, 
one obtains $\log_{10} [E_{\rm jet}/{\rm erg}] = 50.30^{+0.84}_{-0.57}$. 
We use this result as an input for Eq.~\eqref{eq:EGRB}
and sample over the final value employing a 
Gaussian distribution with a width identical to 
the stated uncertainty in~\cite{vanEerten:2018vgj}.

The left panel of Fig.~\ref{fig:GRB} shows our results.
We find that $\log_{10} 
\varepsilon \approx -2$, i.e., 
$1\%$ of the disk rest-mass is converted into 
GRB energy. This generally agrees with existing theoretical studies~\citep{Lee:2007js,Giacomazzo:2012zt} and  
increases the confidence in our GRB analysis. 
Furthermore, we find that the $\tilde{\Lambda}$ 
estimate and the 
constraint on the mass ratio shifts to slightly larger 
values than studying purely AT2017gfo.

\textbf{II. The GRB model of \cite{Wu:2018bxg}:}
An alternative description of GRB170817A is presented in \cite{Wu:2018bxg}. 
They employ the analytic two-parameter ``boosted fireball'' model. 
The model consists of a variety of outflows varying smoothly between a 
highly collimated ultra-relativistic jet and an isotropic fireball. 
Developing a synthetic light curve generator, 
they fit the observational data by performing a 
Markov-Chain Monte Carlo (MCMC) analysis. 
Similar to~\cite{vanEerten:2018vgj}, \cite{Wu:2018bxg} favor a
relativistic structured jet. The jet opening angle is $\sim 5$ degrees 
seen from a viewing angle of $\sim 27$ degree. 

The middle panel of Fig.~\ref{fig:GRB} shows our results.
The estimated GRB energy is $\log10_{E_{\rm jet,50}} = -0.81^{+0.26}_{-0.39}$,
i.e., more than an order of magnitude below the estimated GRB energy of \cite{vanEerten:2018vgj}. 
Consequently,  the estimated value of $\varepsilon$
is smaller. Nevertheless, the constraints on the binary properties and the 
EOS constraints are in agreement between \cite{vanEerten:2018vgj} and \cite{Wu:2018bxg}, i.e.,
the constraints are robust to the systematic difference in energy estimates.

\textbf{III. GRB due to the Blanford-Znajek mechanism~\citep{Wang:2018nye}:}

As a final way to interpret the observed GRB, 
we follow~\citep{Wang:2018nye}. 
In this model, the energy to launch 
the GRB, assuming neutrino annihilation as the central engine, requires 
massive disks masses of the order of $\sim 0.3 M_\odot$. 
Such massive disks are in tension with state-of-the-art numerical 
relativity simulations and disfavor neutrino annihilation as the mechanism 
responsible for the jet-launch. 
On the other hand, magnetic energy extraction requires disk masses about 
one order of magnitude smaller and therefore could act as the central 
engine for GRB170817A. 
Following the discussion of~\cite{Wang:2018nye}, the disk mass necessary 
to explain the observation of GRB170817A based on the Blanford-Znajek (BZ)
mechanism is given by 
\begin{equation}
 m_{\rm disk}^{\rm BZ} = \underbrace{0.0132 M_{\rm \odot} \frac{1}{\mathcal{F}_{\rm GRB}} 
 \frac{E_{\rm GRB}}{10^{51}\rm erg}}_{\sigma} \left( \frac{1 + \sqrt{1-\chi_{\rm BH}^2}}{\chi_{\rm BH}} \right)^2.
 \label{eq:BZ}
\end{equation}
In contrast to~\cite{Wang:2018nye}, we substitute $\frac{0.0132M_\odot}{\mathcal{F}_{\rm GRB}} 
 \frac{E_{\rm GRB}}{10^{51}\rm erg}$ by a single parameter $\sigma$ for which we assume 
 a flat prior with $\log_{10}(\sigma) \in [-4,-1]$. 
Furthermore, we extend the analysis of~\cite{Wang:2018nye}, 
 who used a flat distribution for the BH spin within 
$\chi \in [0.6,0.8]$, by employing Eq.~\eqref{eq:chi_fit} 
to estimate the BH spin, and we substitute the 
disk mass fits of~\cite{Radice:2018pdn} by Eq.~\eqref{eq:mdisk_fit}. 
As in the previous discussions, 
we also incorporate the disk wind ejecta 
via Eq.~\eqref{eq:EGRB}. 
The right panel of Fig.~\ref{fig:GRB} shows our results.
The final results on the tidal deformability, mass ratio, and $M_{\rm TOV}$
are similar to the previous results. 

Very recently, \cite{Wang:2018nye} provided constraints on the EOS 
obtained from a new interpretation of the GRB and its afterglow phase, quoting a constraint of 
$273 < \tilde{\Lambda} < 602$.
Our tests show that this constraint is highly dependent on the particular 
choice of $\sigma$ made in~\cite{Wang:2018nye}. 
Assuming flat priors on all unknown parameters in 
Eq.~\eqref{eq:BZ} creates a prior peaking at $\sim 10^{-2}$. 
This prior choice is the driving mechanism for the very tight constraint 
presented in~\cite{Wang:2018nye} and seems in our opinion 
to be an artifact of the sampling rather than a physical observation. 

\section{Fits to numerical relativity} 

\begin{figure*}
    \centering
    \includegraphics[width=3.4in]{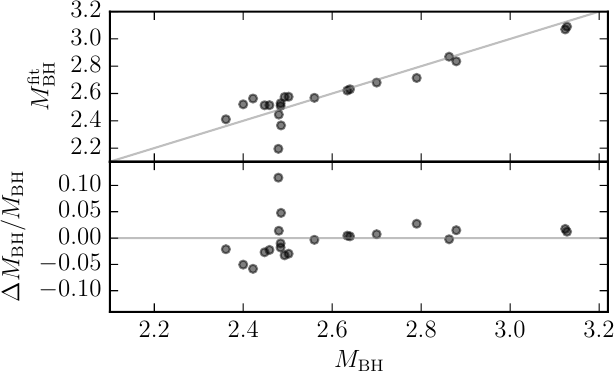}
    \includegraphics[width=3.4in]{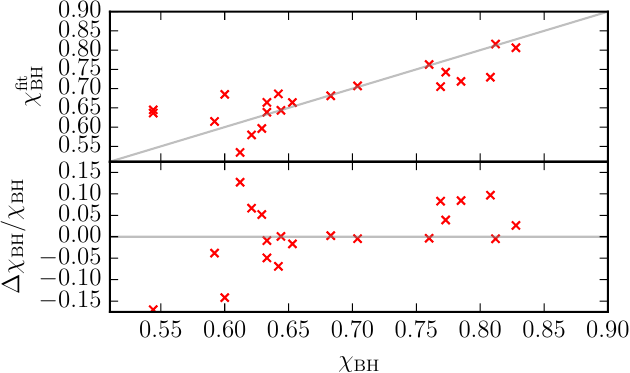}    
    \caption{Quality assessment of the phenomenological descriptions of the 
    BH mass (left panels) and BH spin (right panels). 
    The BH quantities are extracted for BNS simulations of the CoRe catalog, 
    a detailed list is presented in Table~\ref{tab:BH_rem}.}
    \label{fig:BH_fit}
\end{figure*}

\begin{table*}
  \centering
  \caption{
  Overview about the employed numerical relativity dataset to derive the BH remnant fits. 
  Further details are available at \texttt{http://www.computational-relativity.org}. 
  Note that the simulations using data of \protect\cite{Bernuzzi:2015opx} 
  are not identical 
  to the simulations for the same physical configuration available in the CoRe database. 
  The columns refer to: the name of the simulation as stated in the CoRe catalog, 
  the EOS, the total mass $M$, the mass ratio $q$, the tidal deformability $\tilde{\Lambda}$, 
  the references for the simulation and numerical relativity data, 
  the measured BH mass $M_{\rm BH}$ and spin $\chi_{\rm BH}$.}
  \begin{tabular}{l|ccccc|cc}
     Name & EOS & $M$  & $q$& $\tilde{\Lambda}$ & Reference & $M_{\rm BH}$ & $\chi_{\rm BH}$\\
     \hline
BAM:0001:R01 & 2B   & 2.70 & 1.00 & 127 & \citep{Bernuzzi:2014owa} & 2.634 & 0.785\\
BAM:0004:R02 & ALF2 & 2.70 & 1.00 & 730 & \citep{Dietrich:2015iva} & 2.459 & 0.633\\
BAM:0005:R01 & ALF2 & 2.75 & 1.00 & 658 & \citep{Dietrich:2016hky} & 2.493 & 0.653\\
BAM:0011:R01 & ALF2 & 3.00 & 1.00 & 383 & \citep{Dietrich:2018phi} & 2.863 & 0.760\\
BAM:0012:R01 & ALF2 & 2.75 & 1.25 & 671 & \citep{Dietrich:2016hky} & 2.484 & 0.644\\
BAM:0016:R01 & ALF2 & 3.20 & 1.00 & 246 & \citep{Dietrich:2018phi} & 3.128 & 0.812\\
BAM:0017:R01 & ALF2 & 2.75 & 1.50 & 698 & \citep{Dietrich:2016hky} & 2.485 & 0.629\\
BAM:0021:R01 & ALF2 & 2.75 & 1.75 & 731 & \citep{Dietrich:2016hky} & 2.479 & 0.612\\
BAM:0036:R01 & H4   & 2.70 & 1.00 & 1106  & \citep{Dietrich:2015iva} & 2.480& 0.621\\
BAM:0042:R01 & H4   & 2.75 & 1.00 & 993 & \citep{Dietrich:2018phi} & 2.448 & 0.592\\
BAM:0047:R01 & H4   & 3.00 & 1.00 & 567 & \citep{Dietrich:2018phi} & 2.879 & 0.773\\
BAM:0052:R01 & H4   & 3.20 & 1.00 & 359 & \citep{Dietrich:2018phi} & 3.124 & 0.828\\
BAM:0103:R01 & SLy  & 2.70 & 1.00 & 388 & \citep{Dietrich:2015iva} & 2.484 & 0.633\\
BAM:0120:R01 & SLy  & 2.75 & 1.00 & 346 & \citep{Dietrich:2017feu} & 2.640 & 0.769\\
BAM:0123:R02 & SLy  & 2.70 & 1.16 & 490 & \citep{Dietrich:2015iva} & 2.502 & 0.642\\
BAM:0126:R02 & SLy  & 2.75 & 1.25 & 365 & \citep{Dietrich:2016hky} & 2.422 & 0.600\\
BAM:0128:R01 & SLy  & 2.75 & 1.50 & 407 & \citep{Dietrich:2016hky} & 2.361 & 0.544\\
- & LS220 & 2.70 & 1.00 & 684 & \citep{Bernuzzi:2015opx} & 2.40 &  0.544 \\
- & LS220 & 2.83 & 1.04 & 499 & \citep{Bernuzzi:2015opx}  & 2.70 &  0.704 \\
- & SFHo & 2.70 & 1.00 & 422 & \citep{Bernuzzi:2015opx} & 2.56 & 0.683 \\ 
- & SFHo & 2.83 & 1.04 & 312 &  \citep{Bernuzzi:2015opx} & 2.79 &  0.808 \\ 
  \end{tabular}
 \label{tab:BH_rem}
\end{table*}

We now present the fits to numerical relativity we performed which 
are required for our analyses.

\textbf{I. Disk mass}
A crucial ingredient for the analysis in this work and also the recent works 
of \cite{Radice:2018ozg,Wang:2018nye} is the estimate of the debris disk mass $m_{\rm disk}$. 
Here, we revisit the derivation of the phenomenological fit presented
in~\cite{Radice:2018pdn} and employ their fiducial dataset to derive 
an updated version of the fit. 
Figure~15 of~\cite{Radice:2018pdn} shows a clear correlation between $m_{\rm disk}$ and 
the tidal deformability of the binary. 
We suggest that the reason for this clear and prominent correlation 
is related to the limited sample of only four EOSs
in comparison to the wide range of sampled masses,
and the fact that the tidal deformability depends 
strongly on the total mass of the binary, 
$\tilde{\Lambda} \sim \left( M_{\rm tot} / R \right)^{-6}$ ~\citep{De:2018uhw}. 
Already from Fig.~15 of~\cite{Radice:2018pdn}, one sees that setups with the same EOS 
(and hence roughly same radii) but different NS masses, lead to different disk masses.

Here we propose an alternative explanation which naturally accounts for 
the observed phenomenology and scaling with $M_{\rm tot}$.
Merger simulations suggest that the disk mass is accumulated primarily 
through redistribution of matter in the post-merger remnant. 
Thus, the remnant lifetime prior to collapse is found to strongly 
correlate with the amount of disk mass~\citep{Radice:2018xqa}.
Here we suggest that this lifetime is governed to a large degree by 
$M_{\rm tot}/M_{\rm thr}$, where $M_{\rm thr}$ is the threshold mass 
above which the merger undergoes a prompt-collapse (on dynamical timescales).
Thus $M_{\rm tot}/M_{\rm thr}$ is a measure of the stability of 
the post-merger remnant, and following the arguments above, should correlate with $m_{\rm disk}$.

We show in Fig.~2 in the main text the correlation between the 
disk mass and the threshold mass for prompt BH formation, 
where we estimate the prompt collapse threshold as~\cite{Bauswein:2013jpa}:
\begin{equation} 
\label{eq:Mthr}
 M_{\rm thr} = \left(2.38 - 3.606 \frac{M_{\rm TOV}}{R_{1.6M_\odot}}\right)M_{\rm TOV}.
\end{equation}
$M_{\rm TOV}$ denotes the maximum mass of a non-rotating (TOV) NS 
for a given EOS and $R_{1.6M_\odot}$ is the radius of a $1.6M_\odot$ star.  
With the help of Fig.~2 in the main text, 
it becomes clear that the reduction of the 
disk mass relates to the stability of the merger remnant and consequently, 
the disk mass drops abruptly when $M \approx M_{\rm th}$ and 
the remnant undergoes a prompt-collapse. 
This naturally explains the location of the turnover in $m_{\rm disk}$. 

Based on these observations and the fact that the NS radius 
can be related to the tidal deformability by $R = \mathcal{M} (\tilde{\Lambda}/a)^{1/6}$
(with the chirp mass $\mathcal{M}$)~\citep{De:2018uhw}, we conclude that 
the disk mass is a function of the tidal deformability, 
the total mass of the system, 
and the maximum TOV mass $M_{\rm TOV}$.
Contrary to \cite{Kiuchi:2019lls}, 
we do not find a strong dependence on the mass ratio and 
neglect mass ratio effects,
when we analysis some of the numerical relativity 
data of~\cite{Dietrich:2016hky}. This emphasizes a 
thoughtful follow-up study using (if possible) 
different numerical relativity codes to understand the exact 
dependence of the ejecta and disk mass on the mass ratio of the binary.
Therefore, information about the 
densest part of the EOS, encoded in $M_{\rm TOV}$, and the
information at smaller densities, encoded in $\tilde{\Lambda}$ or $R_{1.6M\odot}$, 
are essential for a reliable description of the disk mass. 

To include the dependence of $M_{\rm TOV}$, 
we also find that fitting $\log_{10} (m_{\rm disk})$
instead of $m_{\rm disk}$ leads to a significant 
reduction of the fractional error 
$(m_{\rm disk}-m_{\rm disk}^{\rm fit})/m_{\rm disk}$.
We choose the following functional form
\begin{align}
\log_{10} &\left( m_{\rm disk} \left[ M_{\rm tot}/M_{\rm thr} \right]\right) = \nonumber \\ 
& \max \left(-3,\ a  \left(1 + b \tanh \left[ \frac{c - M_{\rm tot}/M_{\rm thr}}{d} \right] \right) \right),
\label{eq:mdisk_fit}
\end{align}
with $M_{\rm thr} ( M_{\rm TOV}, R_{1.6 {M_\odot}} )$ given by Eq.~(\ref{eq:Mthr}). 
We emphasize that the choice of the exact form of Eq.~(\ref{eq:mdisk_fit}) 
is arbitrary and other expressions are possible. 
The free fitting parameters of Eq.~\eqref{eq:mdisk_fit} 
are 
$a = -31.335$, $b=-0.9760$, $c=1.0474$, $d=0.05957$. 

The mean absolute error of $m_{\rm disk}$ with respect to the original numerical 
relativity data is $0.019M_{\odot}$, 
and we obtain a fractional error of $198\%$ in $m_{\rm disk}$; 
for comparison, the original fit presented in~\cite{Radice:2018pdn} has 
absolute errors of $0.022M_\odot$ and average fractional errors of $749\%$. 
The large fractional error is caused by a small number of setups 
with very small disk masses. 
Fitting the logarithm of the disk mass improves the fit in this region 
of the parameter space, as already discussed in~\cite{Coughlin:2018miv}
for the dynamical ejecta.
We present the fit as a function of the 
$M_{\rm tot}/M_{\rm thr}$ in Fig.~2 in the main text and also present 
the absolute (middle panel) and fractional errors (bottom panel) 
for Eq.~\eqref{eq:mdisk_fit} in comparison with the results of~\cite{Radice:2018pdn}. We point out that in a region around the turning 
point into prompt collapse scenarios, the absolute and fractional errors 
of the new fit are noticeable smaller than in the original version. 

Finally, since we want to relate information extracted from 
the disk mass estimates with the GW measurement, 
we propose to relate the NS radius to the tidal 
deformability via~\cite{De:2018uhw}
\begin{equation}
R_{1.6 {M_\odot}} \simeq \mathcal{M} \left( \frac{\tilde{\Lambda}}{0.0042} \right)^{1/6}. \label{eq:Deetal.}
\end{equation}
While informing this relation adds an additional uncertainty, we find that the fitting residuals 
increase simply to $0.020M_\odot$ and $210\%$. 

\textbf{II. Dynamical Ejecta}
We approximate the mass of the dynamical ejecta by 
\begin{small}
\begin{equation}
\log_{10} m_{\rm dyn}^{\rm fit} = 
\left[ a \frac{(1-2\ 	 C_1) M_1}{C_1}+b\ M_2 \left( \frac{M_1}{M_2} \right)^n +\frac{d}{2} \right] + 
[1 \leftrightarrow 2],
\label{eq:mdyn}
\end{equation}
\end{small}
with $a=-0.0719$, $b=0.2116$, $d=-2.42$, and $n=-2.905$ 
and $C_{1,2}$ denoting the compactnesses of the individual stars. 
The absolute uncertainty of the fit, i.e., $m_{\rm dyn}-m_{\rm dyn}^{\rm fit}$ is $7\times10^{-3}M_\odot$.
Furthermore, we note that while the fractional error of 
$\log_{10} m_{\rm dyn}$ is only $36\%$, the fractional error 
with respect to $m_{\rm dyn}$ is $287\%$ caused by datapoints with very small ejecta 
masses ($\sim 10^{-4} - 10^{-5} M_\odot$). 

The velocity of the dynamical ejecta is given by
\begin{equation}
v_{\rm dyn}^{\rm fit} = \left[a (1 + c\ C_1) \frac{M_1}{M_2} + \frac{b}{2} \right] + [ 1 \leftrightarrow 2],
\label{eq:vdyn}
\end{equation}
where $a=-0.3090$, $b=0.657$, and $c=-1.879$. 
The average absolute error of the fit is $\Delta v_{\rm dyn} = 0.03$ 
and the fractional error is $18\%$.

\textbf{III. BH properties}
The large set of numerical relativity data publicly 
released in the CoRe catalog~\citep{Dietrich:2018phi} 
together with results published in~\cite{Bernuzzi:2015opx}
allows us to derive phenomenological fits 
for the BH mass and spin. 
A detailed list of the employed simulations and the BH properties
is presented in Tab.~\ref{tab:BH_rem}.
We restrict our consideration to non-spinning NSs, but plan to extend 
the presented results in the future once a larger set 
of spinning BNS configurations is available. 
Furthermore, we consider only cases for which an almost stationary 
state is reached after BH formation, so that remnant properties can 
be extracted reliably. 
Thus, we do not consider setups for which the BH mass increases significantly 
due to accretion or for which the black hole 
mass decreases due to insufficient resolution. 

Trivially, we find that with an increasing total mass, 
the final black hole mass and angular momentum increases almost linearly. 
For unequal mass mergers, $M_{\rm BH}$ 
and $\chi_{\rm BH}$ decrease. 
Based on this observation, we propose a functional dependence of 
$M_{\rm BH}  \propto \nu^\alpha$ (where $\nu$ refers to the symmetric mass ratio). 
The coefficient $\alpha$ is chosen to be two, 
which is motivated by predictions for BBH systems~\citep{Healy:2017mvh}.
Considering the imprint of the EOS, we find that for larger 
values of $\tilde{\Lambda}$, the final black hole mass decreases, 
which follows from the observation that the 
disk mass increases with $\tilde{\Lambda}$. 
As a simple ansatz, we choose: 
\begin{equation}
 M_{\rm BH} = a \left(\frac{\nu}{0.25}\right)^2 \left(M+b \ \frac{\tilde{\Lambda}}{400}\right) \label{eq:MBH_fit}
\end{equation}
with $a=0.980$ and $b=-0.093$.
The mean average absolute error of the fit is $0.065M_\odot$ 
and the fractional error is $2.6\%$. 

We find in our dataset that the BH mass and spin are 
strongly correlated. This motivates the use of a similar 
functional behavior for the BH spin as for the BH mass. 
However, we extend Eq.~\eqref{eq:MBH_fit} by 
(i) adding an additional constant, and 
(ii) incorporating the fact that the dimensionless spin 
is restricted to be $\chi \leq 1$.
The final fitting function is 
\begin{equation}
 \chi_{\rm BH} = \tanh \left[ a \nu^2 (M+b\ \tilde{\Lambda}) +c \right] \label{eq:chi_fit} 
\end{equation}
with $a=0.537$, $b=-0.185$, and $c=-0.514$.
The mean average absolute error of the fit is $0.039$ 
and the fractional error is $6.1\%$. 
The remnant property dataset and the 
corresponding fit results are presented 
in Fig.~\ref{fig:BH_fit}.


\bsp	
\label{lastpage}
\end{document}